\documentclass[11pt]{article}

\usepackage{geometry}   
\usepackage{authblk}

\usepackage{multirow}

\usepackage[utf8]{inputenc} 
\usepackage{hyperref}       
\usepackage{url}            
\usepackage{booktabs}       
\usepackage{amsfonts}       
\usepackage{nicefrac}       
\usepackage{microtype}      

\usepackage{amsmath,amsfonts,amssymb,amsthm,array}

\usepackage{algorithm}
\usepackage{caption}
\usepackage{subcaption}
\usepackage{algpseudocode}
\usepackage{mdframed} 
\usepackage{thmtools}

\usepackage{graphicx} 
\usepackage{textcomp}

\newcommand{\R}{\mathbb{R}}

\def\<#1,#2>{\langle #1,#2\rangle}

\usepackage[colorinlistoftodos,bordercolor=orange,backgroundcolor=orange!20,linecolor=orange,textsize=scriptsize]{todonotes}

\newcommand{\eqdef}{\; { := }\;}

\usepackage{natbib}
\setcitestyle{authoryear,round}

\title{\bf Deep Learning for Multi-Country GDP Prediction: A Study of Model Performance and Data Impact   }

\date{September 4, 2024}

\author[1]{Huaqing Xie}
\author[1]{Xingcheng Xu}
\author[1]{Fangjia Yan}
\author[1]{Xun Qian\textsuperscript{*}}
\author[1,2]{Yanqing Yang\thanks{Corresponding author.}}

\affil[1]{Shanghai Artificial Intelligence Laboratory}
\affil[2]{Fudan University}
\affil[ ]{xiehuaqing@pjlab.org.cn, xingcheng.xu18@gmail.com, yanfangjia@pjlab.org.cn, qianxun@pjlab.org.cn, yanqingyang@fudan.edu.cn}

\begin{document}
	
	\maketitle
	
\begin{abstract}
GDP is a vital measure of a country's economic health, reflecting the total value of goods and services produced. Forecasting GDP growth is essential for economic planning, as it helps governments, businesses, and investors anticipate trends, make informed decisions, and promote stability and growth. While most previous works focus on the prediction of the GDP growth rate for a single country or by machine learning methods, in this paper we give a comprehensive study on the GDP growth forecasting in the multi-country scenario by deep learning algorithms. For the prediction of the GDP growth where only GDP growth values are used, linear regression is generally better than deep learning algorithms. However, for the regression and the prediction of the GDP growth with selected economic indicators, deep learning algorithms could be superior to linear regression. We also investigate the influence of the novel data -- the light intensity data on the prediction of the GDP growth, and numerical experiments indicate that they do not necessarily improve the prediction performance. Code is provided at https://github.com/Sariel2018/Multi-Country-GDP-Prediction.git. 
\end{abstract}

\section{Introduction}

Gross Domestic Product (GDP) is a critical measure of a country's economic health, reflecting the total value of all goods and services produced over a specific period. It serves as a comprehensive indicator of a nation's economic activity, influencing government policy, investment decisions, and international comparisons. GDP is a key metric for assessing economic performance, guiding fiscal and monetary policy, and shaping economic development strategies.

Forecasting GDP growth rate is crucial for economic planning and decision-making. It provides valuable insights into the future direction of an economy, helping governments, businesses, and investors make informed decisions. Accurate GDP growth predictions allow policymakers to anticipate economic trends, adjust fiscal and monetary policies accordingly, and implement measures to promote stability and growth. For businesses, understanding expected economic conditions helps in strategic planning, resource allocation, and risk management. Investors rely on GDP growth forecasts to assess market conditions and make investment choices. Thus, GDP growth rate predictions are essential for shaping economic strategies and fostering sustainable development.

Machine learning has advanced GDP forecasting by building on Dynamic Factor Models (DFMs), first applied by \cite{giannone2008nowcasting} for real-time GDP forecasting, influencing Federal Reserve policy. Central banks like the ECB \citep{marozzi2021ecb} and institutions such as the World Bank \citep{dauphin2022nowcasting} have since adopted similar models.

The integration of machine learning has further enhanced forecasting accuracy. \cite{richardson2021nowcasting} showed that a machine learning model using over 600 variables outperformed traditional AR and DFM models in forecasting New Zealand's GDP, particularly during the COVID-19 pandemic. \cite{woloszko2020tracking} also found that machine learning models using Google data significantly improved GDP growth rate forecasts for OECD countries. \cite{yang2024machine} highlighted these benefits by applying machine learning models like gradient boosting trees, random forests, and kernel ridge regression to forecast China's GDP, demonstrating their superior performance over traditional methods.

There are many works on the prediction of GDP growth by deep learning methods as well. In \citep{widjaja2023predicting}, four key economic indicators were utilized to predict Indonesia's quarterly GDP growth using Multiple Linear Regression (MLR), K-Nearest Neighbours (K-NN), and Artificial Neural Network (ANN) models, with MLR outperforming the others based on RMSE values. Similarly, \cite{jahn2018artificial} employed ANN regression models to forecast GDP growth rates of 15 industrialized economies between 1996 and 2016, demonstrating that ANN provided more accurate and flexible predictions compared to linear models, especially in capturing time trends. In \citep{da2020systematic}, various models, including SARIMA, Holt-Winters, dynamic linear models, and neural networks, were compared for Brazilian GDP forecasting. The Multilayer Perceptron (MLP) outperformed others in both in-sample and out-sample predictions, effectively capturing GDP growth rates.

To address both linear and nonlinear components in GDP forecasting, \cite{chaudharyhybrids} introduced hybrid models combining ARIMA and ANN techniques for Nepal's GDP prediction, where the hybrid models achieved superior accuracy over standalone approaches. Focusing on deep learning methods, \cite{sa2020prediction} applied Long Short-Term Memory (LSTM) and Recurrent Neural Network (RNN) architectures to predict Indonesia's GDP fluctuations during the COVID-19 pandemic, achieving accuracy rates between $80\%$ and $90\%,$ thus highlighting their effectiveness in handling sudden economic changes.

In \citep{tuncsipermodelling}, autoregressive deep learning networks were developed to model Türkiye's GDP and per capita income from 1960 to 2021 using past values of the target variables, resulting in high accuracy as evidenced by performance metrics like $R^2$, MAE, MAPE, and RMSE. Exploring machine learning algorithms further, \cite{premraj2019forecasting} conducted a comprehensive comparison between models such as BART, GLMNET, GBM, and XGBoost against traditional time series methods like ARIMA and VAR across multiple economies, concluding that multivariate VAR models were generally superior, though XGBoost provided valuable insights by emphasizing different influential variables.

For scenarios with limited data, \cite{priambodo2019predicting} demonstrated that K-Nearest Neighbour regression could effectively predict Indonesia's GDP during the 1998 economic crisis, outperforming both backpropagation neural networks and MLR models. Addressing variable selection, \cite{malik2021time} utilized stepwise regression techniques on 14 sub-sectors of Pakistan's economy to construct an appropriate time series model for GDP growth prediction, resulting in a model with nine significant predictors validated through various diagnostic tests.

\cite{jena2021impact} applied a multi-layer ANN model to forecast the COVID-19 pandemic's impact on the GDP of eight major economies for the April–June 2020 quarter, achieving forecasting errors of less than $2\%$ and revealing substantial GDP declines necessitating urgent policy responses. In \citep{ortega2021multimodal}, a novel multimodal approach combined historical GDP data with Twitter activity using a two-stage architecture involving a multi-task autoencoder and a multimodal network, providing timely and accurate regional GDP predictions in Spain and effectively capturing the economic effects of the COVID-19 pandemic.

Among these works, most of them aimed to predict the GDP growth rates of one single country. For the multi-country case, \cite{premraj2019forecasting}, \cite{woloszko2020tracking}, and \cite{dauphin2022nowcasting} adopted machine learning algorithms, while \cite{jahn2018artificial} and \cite{jena2021impact} predicted the GDP growth by MLP, but only previous GDP growth rates were used there.

There are also some works on the time series forecasting. Some of them are the variants of the Transformer model, such as Informer \citep{zhou2021informer}, Autoformer \citep{wu2021autoformer}, FEDformer \citep{zhou2022fedformer}, PatchTST \citep{nie2022time} et al. There are also some time series forecasting works based on large language models (LLMs). For instance, LLMTime \citep{gruver2024large}, Time-LLM \citep{jin2023time}, GPT4TS \citep{zhou2023one}. There are also some foundation models for time series forecasting which are pretrained on a large corpus of time series data, such as Lag-Llama \citep{rasul2024lag}, TimesFM \citep{das2023decoder}, Chronos \citep{ansari2024chronos}, and UniTS \citep{gao2024units}. 

In this paper, we study the prediction of GDP growth in the multi-country regime by deep learning algorithms with multiple economic indicators and novel data. We propose the representation transformer model for the regression of the GDP growth which is a combination of representation in LLM and transformer. For the regression of the annually GDP growth, MLP has
comparable performance with linear regression, but for the regression of the quarterly GDP growth, MLP is better than the linear regression. For the regression of the annually GDP growth, representation transformer is worse than MLP. However, representation transformer can deal with the case where the number of economic indicators of data is variable, while MLP and linear regression could not. 

For the prediction of the quarterly GDP growth by autoregression where only GDP growth values are used, with filtered data by some selected economic indicators, LSTM has comparable performance with linear regression. But  linear regression is better than LSTM for all GDP growth data. TimeFM,Time-LLM, and PatchTST are generally worse than linear regression. 

For the prediction of the quarterly GDP growth with multi-indicator data, LSTM is generally better than linear regression. Time-LLM and PatchTST are actually comparable and are both better than LSTM generally, but they actually could not characterize the impacts of these economic indicators during the inference time. 

In all cases, the light intensity data do not necessarily improve the prediction performance.

\section{Problem Formulation}

Denote the economic indicator at time $t$ as $x_i^t \in \R$, and the GDP growth at time $t$ as $y^t$. There are two regimes for the prediction of GDP growth rate. One is using $\{ x_i^t \}_{i=1}^n$ to predict $y^t$, which is also called regression. The other is predicting the GDP growth $y^t$ by previous economic indicators $x_i^k$ and/or previous GDP growth $y^k$ for $k<t$. 

In this paper, we consider these two regimes respectively. In Section  \ref{sec:regression}, we study the regression of both annually and quarterly GDP growth. In Section \ref{sec:pregdp}, we study the prediction of the quarterly GDP growth $y^t$ by using $(y^{t-h}, ..., y^{t-1})$, where $h$ is the sequence length. Denote 
\begin{equation}\label{eq:zt}
z^t \eqdef (x_1^t, ..., x_n^t, y^t)^\top \in \R^{n+1}.
\end{equation}
In Section \ref{sec:premulgdp}, we study the prediction of the quarterly GDP growth $y^t$ by utilizing $(z^{t-h}, ..., z^{t-1})$. In all these cases, we also study the influence of the novel data -- the light intensity. The nighttime lights data is derived from monthly night-time remote sensing images captured by the Visible Infrared Imaging Radiometer Suite (VIIRS) onboard the NPP satellite. This data has a spatial resolution of 0.004 degrees and covers the period from the launch of the NPP satellite in 2012 to the present. The data has undergone stray light correction, which allows for an accurate representation of temporal changes in local lighting brightness. Nighttime lights data encompasses comprehensive information such as population, transportation, and economic development levels, making it suitable for various development economics research. The raw data is stored in the widely used GeoTIFF format for geographic information and is converted to brightness values using the Rasterio library of Python.

For all the experiments, we divide the train and test sets by time. The test set for the period $2013--2019$ consists of the data in the last year, and for the other periods consists of the data of the last two years. For all the deep learning algorithms, we use k-fold cross-validation method and get two checkpoints. One checkpoint is the best validation checkpoint, which is the best of the $k$ checkpoints with the best average validation loss. The other checkpoint is the checkpoint which is trained on the full train set with all the hyperparameters used by the best validation checkpoint. We use $k=5$ in this paper. For all the deep learning algorithms, we search the best hyperparameters by grid search. For all models, the input data are normalized by the minimum and maximum values of the train and test sets.

\section{Regression for the GDP growth}\label{sec:regression}

In this section, we study the regression for the annually and quarterly GDP growth, i.e., predicting $y^t$ by $(x_1^t, ..., x_n^t)$. We first describe the data we used. 

Based on the 2023 GDP rankings from the World Bank, we selected 21 countries: the United States, China, Germany, Japan, India, the United Kingdom, France, Brazil, Italy, Canada, Russia, Mexico, Australia, South Korea, Spain, Indonesia, the Netherlands, Turkey, Saudi Arabia, Switzerland, and Poland.

We manually selected 13 annual economic indicators from the WEO database of the IMF. These indicators are: `Rural population growth (annual $\%$)', `General government final consumption expenditure (annual $\%$ growth)', `Consumer price index (2010 = 100)', `Exports of goods and services (annual $\%$ growth)', `Urban population growth (annual $\%$)', `GDP growth (annual $\%$)', `Population growth (annual $\%$)', `Inflation, GDP deflator (annual $\%$)', `Imports of goods and services (annual $\%$ growth)', `Final consumption expenditure (annual $\%$ growth)', `Unemployment, total ($\%$ of total labor force) (national estimate)', `Inflation, consumer prices (annual $\%$)', `Gross fixed capital formation (annual $\%$ growth)' and `Households and NPISHs Final consumption expenditure (annual $\%$ growth)'.

For the quarterly data, we selected 20 economic indicators, including `Export Value', `Industrial Added Value', `Stock Market Capitalization', `Balance of Payments - Financial Account Balance', `Balance of Payments - Current Account Balance', `Balance of Payments - Current Account Credit', `Balance of Payments - Current Account Debit', `Balance of Payments - Capital Account Balance', `Balance of Payments - Capital Account Credit', `Balance of Payments - Capital Account Debit', `Overall Balance of Payments', `International Investment Position - Assets', `International Investment Position - Liabilities', `Net International Investment Position', `Import Value', `Nominal Effective Exchange Rate', `Retail Sales', `CPI (Consumer Price Index)', `Unemployment Rate' and `Central Bank Policy Rate'. The data was sourced from the financial institution WIND, with the original sources traceable to the World Bank, IMF, and national statistical offices.

We use there models for the regression of the GDP growth rate, i.e., the linear regression, MLP, and representation transformer(RT). We introduce the representation transformer model next.

\paragraph{Representation Transformer.}
Large language models have significant influence in the field of artificial intelligence since the emergence of ChatGPT \citep{brown2020language}. LLMs can provide good representations for the text. For the limited data in the prediction of the GDP growth, the information contained in the meaning of variable names could be used by the adoption of LLMs. Hence, we first write a description for every selected economic indicator, and then input the text description into a LLM to get a representation vector. Since the representation by LLMs is usually a high dimensional vector, we use a projection layer to reduce the dimension. Then we also concatenate the value of the corresponding variable with multiple times to the projected vector. At last, we use transformer \citep{vaswani2017attention} to obtain the predicted GDP growth. The detailed neural network architecture of RT is described as follows. 

For each economic indicator $x_i^t$, we write a text description $Text_i^t$ and input it to the InternVL-Chat-V1-5\footnote{https://huggingface.co/OpenGVLab/InternVL-Chat-V1-5. We use VLM since it can handle images as well in case we may need to deal with vision data.} model, whose dimension of representation vectors is 6144. For instance, the description for `Exports of goods and services' is: The index ``Volume of exports of goods and services" with the unit ``Percent change" measures the percentage change in the quantity of goods and services that a country sells to other nations over a specific period, usually a year, compared to the previous period. In this year, the index is $\{value\}$ percent change. The ``$\{value\}$" in the description is replaced by $x_i^t$. The descriptions of all the economic indicators are in the Appendix. We choose the representation vector of the last token in $Text_i^t$ before the last fully-connected layer as $Rep_i^t$. Then, 
\begin{eqnarray*}
v_i^t &=& W_1 Rep_i^t + b_1,\\ 
u_i^t &=& (x_i^t, ..., x_i^t)^\top \in \R^{dim}, \\ 
c_i^t &=& concat(v_i^t, u_i^t) + PositionEmbedding, \\ 
(o_1^t, ..., o_n^t)   &=& TransformerEncoder(c_1^t, ..., c_n^t), \\ 
O^t &=& mean(o_1^t, ..., o_n^t), \\
y^t &=& W_2 O^t + b_2,
\end{eqnarray*}
where $dim$ is a hyperparameter. 

The regression of the GDP growth rates for linear regression, MLP, and RT are in Tables \ref{tab:mlp} and \ref{tab:RT}. For the light column, $\texttimes$ denotes the light intensity data were not used. The ``sum" refers to the total light intensity of all pixels within a specific country or region. The ``mean" indicates the average light intensity, while the ``std" represents the standard deviation of light intensity values. For each time period, we calculate the average values of these light indicators. Therefore, ``every month mean" refers to the average light intensity for every month.

\begin{table}[htbp]
  \centering
  \caption{Regression of the annually and quarterly GDP growth rates results for linear regression and MLP.}
  \resizebox{\textwidth}{!}{%
    \begin{tabular}{>{\centering\arraybackslash}p{1cm}|c|>{\centering\arraybackslash}p{6.46em}|>{\centering\arraybackslash}p{5.585em}|c|ccc|ccc|ccc}
    \toprule
    \multicolumn{1}{>{\centering\arraybackslash}p{1cm}|}{\multirow{2}[4]{*}{Dataset}} & \multicolumn{1}{c|}{\multirow{2}[4]{*}{Period}} & \multirow{2}[4]{*}{Light} & \multirow{2}[4]{*}{Train: Test} & \multicolumn{1}{c|}{\multirow{2}[4]{*}{Dims}} & \multicolumn{3}{>{\centering\arraybackslash}p{12.12em}|}{Linear Regression} & \multicolumn{3}{>{\centering\arraybackslash}p{12.12em}|}{MLP Best Valid Model}& \multicolumn{3}{>{\centering\arraybackslash}p{12.12em}}{MLP Final Model} \\
    \cmidrule{6-14}        
    & & & & & \multicolumn{1}{>{\centering\arraybackslash}p{4.04em}}{MAE} & \multicolumn{1}{>{\centering\arraybackslash}p{4.04em}}{MSE} & \multicolumn{1}{>{\centering\arraybackslash}p{4.04em}|}{RMSE} & \multicolumn{1}{>{\centering\arraybackslash}p{4.04em}}{MAE} & \multicolumn{1}{>{\centering\arraybackslash}p{4.04em}}{MSE} & \multicolumn{1}{>{\centering\arraybackslash}p{4.04em}|}{RMSE} & \multicolumn{1}{>{\centering\arraybackslash}p{4.04em}}{MAE} & \multicolumn{1}{>{\centering\arraybackslash}p{4.04em}}{MSE} & \multicolumn{1}{>{\centering\arraybackslash}p{4.04em}}{RMSE} \\
    \midrule
    \multicolumn{1}{c|}{\multirow{6}[12]{*}{yearly}} & \multicolumn{1}{>{\centering\arraybackslash}p{4.04em}|}{80-07} & \texttimes   & 407: 38 & 13  & 0.7517 & 1.7630 & 1.3278 & 0.6648 & 0.7746 & 0.8801 & 0.6173 & 0.6928 & 0.8323 \\
    \cmidrule{2-14}        
    & \multicolumn{1}{>{\centering\arraybackslash}p{4.04em}|}{80-19} & \texttimes   & 635: 40 & 13  & 0.4177 & 0.3269 & 0.5717 & 0.5494 & 0.5684 & 0.7539 & 0.5818 & 0.5058 & 0.7112 \\
    \cmidrule{2-14}        
    & \multicolumn{1}{c|}{\multirow{4}[8]{*}{13-19}} & \texttimes   & 115: 20 & 13  & 0.3324 & 0.1881 & 0.4337 & 0.3602 & 0.2341 & 0.4839 & 0.3100 & 0.1872 & 0.4326 \\
    \cmidrule{3-3}        
    & & sum\textbackslash{}mean\textbackslash{}std & 115: 20 & 16  & 0.2891 & 0.1489 & 0.3859 & 0.4975 & 0.4516 & 0.6720 & 0.3069 & 0.1534 & 0.3917 \\
    \cmidrule{3-3}        
    & & mean & 115: 20 & 14  & 0.3213 & 0.1697 & 0.4120 & 0.5154 & 0.5936 & 0.7704 & 0.2333 & 0.1179 & 0.3433 \\
    \cmidrule{3-3}        
    & & every month mean & 115: 20 & 25  & 0.3817 & 0.2135 & 0.4620 & 0.6257 & 0.8419 & 0.9175 & 0.3876 & 0.2495 & 0.4995 \\
    \midrule
    \multicolumn{1}{c|}{\multirow{5}[10]{*}{quarterly}} & \multicolumn{1}{>{\centering\arraybackslash}p{4.04em}|}{95-19} & \texttimes   & 787: 123 & 20  & 1.3195 & 5.1222 & 2.2632 & 0.9894 & 1.7844 & 1.3358 & 0.9488 & 1.7157 & 1.3098 \\
    \cmidrule{2-14}        
    & \multicolumn{1}{c|}{\multirow{4}[8]{*}{13-19}} & \texttimes   & 352: 61 & 20  & 1.0894 & 2.7885 & 1.6699 & 0.8334 & 1.0991 & 1.0484 & 0.8699 & 1.1974 & 1.0943 \\
    \cmidrule{3-3}        
    & & sum\textbackslash{}mean\textbackslash{}std & 352: 61 & 23  & 1.1358 & 3.0647 & 1.7506 & 0.7697 & 1.0284 & 1.0141 & 0.7610 & 0.9112 & 0.9546 \\
    \cmidrule{3-3}        
    & & mean & 352: 61 & 21  & 1.0866 & 2.9060 & 1.7047 & 0.7785 & 0.9762 & 0.9880 & 0.8116 & 1.0479 & 1.0237 \\
    \cmidrule{3-3}        
    & & every month mean & 352: 61 & 23  & 1.0928 & 2.9099 & 1.7058 & 0.8260 & 1.1641 & 1.0789 & 0.7606 & 0.9821 & 0.9910 \\
    \bottomrule
    \end{tabular}%
  }
  \label{tab:mlp}
\end{table}

\begin{table}[htbp]
  \centering
  \caption{Regression of the annually GDP growth rates results for RT.}
  \resizebox{\textwidth}{!}{%
    \begin{tabular}{c|c|>{\centering\arraybackslash}p{6.46em}|>{\centering\arraybackslash}p{5.585em}|>{\centering\arraybackslash}p{5.04em}|ccc|ccc}
    \toprule
    \multicolumn{1}{c|}{\multirow{2}[4]{*}{Dataset}} & \multicolumn{1}{c|}{\multirow{2}[4]{*}{Period}} & \multirow{2}[4]{*}{Light} & \multirow{2}[4]{*}{Train: Test} & \multirow{2}[4]{*}{Input Shape} & \multicolumn{3}{>{\centering\arraybackslash}p{12.12em}|}{RT Best Valid Model} & \multicolumn{3}{>{\centering\arraybackslash}p{12.12em}}{RT Final Model} \\
    \cmidrule{6-11}        
    & & & & & \multicolumn{1}{>{\centering\arraybackslash}p{4.04em}}{MAE} & \multicolumn{1}{>{\centering\arraybackslash}p{4.04em}}{MSE} & \multicolumn{1}{>{\centering\arraybackslash}p{4.04em}|}{RMSE} & \multicolumn{1}{>{\centering\arraybackslash}p{4.04em}}{MAE} & \multicolumn{1}{>{\centering\arraybackslash}p{4.04em}}{MSE} & \multicolumn{1}{>{\centering\arraybackslash}p{4.04em}}{RMSE} \\
    \midrule
    \multicolumn{1}{c|}{\multirow{6}[10]{*}{yearly}} & \multicolumn{1}{>{\centering\arraybackslash}p{4.04em}|}{80-07} & \texttimes   & 407: 38 & 13, 6145& 0.8817 & 1.7984 & 1.3411 & 0.8472 & 1.2041 & 1.0973 \\
    \cmidrule{2-11}        
    & \multicolumn{1}{>{\centering\arraybackslash}p{4.04em}|}{80-19} & \texttimes   & 635: 40 & 13, 6145 & 0.5910 & 0.7588 & 0.8711 & 0.6299 & 0.6817 & 0.8257 \\
    \cmidrule{2-11}        
    & \multicolumn{1}{c|}{\multirow{4}[6]{*}{13-19}} & \texttimes   & 115: 20 & 13, 6145 & 0.4693 & 0.4105 & 0.6407 & 0.5143 & 0.5895 & 0.7678 \\
    \cmidrule{3-3}        
    & & \multirow{2}[2]{*}{sum\textbackslash{}mean\textbackslash{}std} & \multirow{2}[2]{*}{115: 20} & \multirow{2}[2]{*}{14, 6145} & \multirow{2}[2]{*}{0.8003} & \multirow{2}[2]{*}{1.1114} & \multirow{2}[2]{*}{1.0542} & \multirow{2}[2]{*}{1.0615} & \multirow{2}[2]{*}{1.8366} & \multirow{2}[2]{*}{1.3552} \\
    & & \multicolumn{1}{c|}{} & \multicolumn{1}{c|}{} & \multicolumn{1}{c|}{} &     &     &     &     &     &  \\
    \cmidrule{3-3}        
    & & every month mean & 115: 20 &  25, 6145 & 0.8246 & 1.1713 & 1.0823 & 0.6106 & 0.7755 & 0.8806 \\
    \bottomrule
    \end{tabular}%
  }
  \label{tab:RT}
\end{table}

From Table \ref{tab:mlp}, we can see that for the regression of the annually GDP growth,  MLP has comparable results with linear regression, but for the regression of the quarterly GDP growth, MLP is better than the linear regression. Furthermore, the usage of the light intensity data do not always make the results better. Table \ref{tab:RT} shows that for the regression of the annually GDP growth, representation transformer is worse than MLP.  However, representation transformer can deal with the case where the number of economic indicators of data is variable, while MLP and linear regression could not. Moreover, the performance of RT could be improved by the development of LLMs, and by finetuning the part at the beginning of the LLM like Time-LLM \citep{jin2023time}, which could be future work.

\section{Prediction of the quarterly GDP growth by autoregression}\label{sec:pregdp}

In this section, we predict the quarterly GDP growth $y^t$ by previous GDP growth rates $(y^{t-h}, ..., y^{t-1})$. For the period between 2013 and 2019, we also add the light intensity dimension to investigate the influence of the light data. We use five models in this section, i.e., linear regression, LSTM, TimesFM \citep{gruver2024large}, Time-LLM\footnote{We use GPT-2 for the LLM in Time-LLM in our experiments.} \citep{jin2023time}, and PatchTST \citep{nie2022time}. For the GDP growth sequence data, we consider two scenarios. The first scenario involves directly using the GDP data from the multidimensional dataset, which allows for comparison with the results obtained using multidimensional data. The second scenario involves using all historical GDP data to capture the cyclical changes and long-term trends inherent in GDP itself. Due to the availability of data from other dimensions, the data used in the first scenario is a subset of the data used in the second scenario. When the light data are used, the data type is actually the same as that in Section \ref{sec:premulgdp}, and the setups are the same as that in Section \ref{sec:premulgdp} as well. For simplicity, we omit the setups here.

\begin{table}[htbp]
  \centering
  \caption{Prediction of the quarterly GDP growth by autoregression results for linear regression and LSTM.}
\resizebox{\textwidth}{!}{%
    \begin{tabular}{c|c|c|c|c|c|ccc|ccc|ccc}
    \toprule
    \multirow{2}[3]{*}{Dataset} & \multicolumn{1}{c|}{\multirow{2}[3]{*}{Period}} & \multicolumn{1}{c|}{\multirow{2}[3]{*}{Light}} & \multicolumn{1}{c|}{\multirow{2}[3]{*}{\parbox{1cm}{\centering Seq Length}}} & \multicolumn{1}{c|}{\multirow{2}[3]{*}{Train: Test}} & \multirow{2}[3]{*}{Dims} & \multicolumn{3}{c|}{Linear Regression} & \multicolumn{3}{c|}{LSTM best Valid model} & \multicolumn{3}{c}{LSTM Final Model} \\
\cmidrule{7-15}        &     &     & \multicolumn{1}{c|}{} & \multicolumn{1}{c|}{} &     & \multicolumn{1}{>{\centering\arraybackslash}p{3.75em}}{MAE} & \multicolumn{1}{>{\centering\arraybackslash}p{5.21em}}{MSE} & \multicolumn{1}{>{\centering\arraybackslash}p{5.29em}|}{RMSE} & \multicolumn{1}{>{\centering\arraybackslash}p{6.54em}}{MAE} & \multicolumn{1}{>{\centering\arraybackslash}p{3.75em}}{MSE} & \multicolumn{1}{>{\centering\arraybackslash}p{3.75em}|}{RMSE} & \multicolumn{1}{>{\centering\arraybackslash}p{3.75em}}{MAE} & \multicolumn{1}{>{\centering\arraybackslash}p{3.75em}}{MSE} & \multicolumn{1}{>{\centering\arraybackslash}p{4.04em}}{RMSE} \\
\midrule
    \multicolumn{1}{c|}{\multirow{15}[9]{*}{\parbox{1.5cm}{\centering GDP Filtered by Other Variables}}} & \multicolumn{1}{c|}{\multirow{3}[1]{*}{95-19}} & \multicolumn{1}{c|}{\multirow{3}[1]{*}{\texttimes }} & 8   & 592: 108 & 1   & 0.4857 & 0.4753 & 0.6894 & 0.4828 & 0.4580 & 0.6768 & 0.5964 & 0.6082 & 0.7799 \\
        &     &     & 10  & 558: 104 & 1   & 0.4592 & 0.4387 & 0.6623 & 0.4699 & 0.4299 & 0.6556 & 0.5344 & 0.4936 & 0.7025 \\
        &     &     & 12  & 530: 100 & 1   & 0.4712 & 0.4555 & 0.6749 & 0.5012 & 0.4864 & 0.6974 & 0.4745 & 0.4293 & 0.6552 \\
\cmidrule{2-15}        & \multicolumn{1}{c|}{\multirow{12}[8]{*}{13-19}} & \multicolumn{1}{c|}{\multirow{3}[2]{*}{\texttimes }} & 8 & 202: 56 & 1   & 0.4882 & 0.4280 & 0.6542 & 0.7148 & 0.9605 & 0.9801 & 0.5101 & 0.4737 & 0.6883 \\
        &     &     & 10 & 170: 56 & 1   & 0.4195 & 0.3725 & 0.6103 & 0.4858 & 0.5223 & 0.7227 & 0.4692 & 0.4907 & 0.7005 \\
        &     &     & 12 & 143: 52 & 1   & 0.4503 & 0.4443 & 0.6665 & 0.6056 & 0.6380 & 0.7987 & 0.5555 & 0.6716 & 0.8195 \\
\cmidrule{3-15}        &     & \multicolumn{1}{c|}{\multirow{3}[2]{*}{\parbox{1cm}{\centering sum\textbackslash{} \\ mean\textbackslash{} \\ std}}} & 8   & 202: 56 & 4   & 0.5209 & 0.5153 & 0.7178 & 0.7816 & 1.3233 & 1.1504 & 0.5233 & 0.5256 & 0.7250 \\
        &     &     & 10  & 170: 56 & 4   & 0.5926 & 0.6827 & 0.8263 & 0.5831 & 0.6635 & 0.8145 & 0.5530 & 0.6443 & 0.8027 \\
        &     &     & 12  & 143: 52 & 4   & 0.6117 & 0.7339 & 0.8567 & 0.7072 & 0.8605 & 0.9276 & 0.6818 & 0.9261 & 0.9624 \\
\cmidrule{3-15}        &     & \multicolumn{1}{c|}{\multirow{3}[2]{*}{mean}} & 8   & 202: 56 & 2   & 0.4776 & 0.4269 & 0.6534 & 0.7807 & 1.1871 & 1.0895 & 0.4789 & 0.4574 & 0.6763 \\
        &     &     & 10  & 170: 56 & 2   & 0.5059 & 0.5275 & 0.7263 & 0.5275 & 0.5587 & 0.7475 & 0.6645 & 0.7913 & 0.8895 \\
        &     &     & 12  & 143: 52 & 2   & 0.5389 & 0.6529 & 0.8080 & 0.6655 & 0.8172 & 0.9040 & 0.4782 & 0.5429 & 0.7368 \\
\cmidrule{3-15}        &     & \multicolumn{1}{c|}{\multirow{3}[2]{*}{\parbox{1cm}{\centering every month mean}}} & 8   & 202: 56 & 4   & 0.5546 & 0.5199 & 0.7210 & 0.6813 & 0.8770 & 0.9365 & 0.4996 & 0.5401 & 0.7349 \\
        &     &     & 10  & 170: 56 & 4   & 0.5750 & 0.6550 & 0.8093 & 1.0508 & 1.9837 & 1.4084 & 0.7039 & 0.9497 & 0.9745 \\
        &     &     & 12  & 143: 52 & 4   & 0.5953 & 0.7487 & 0.8652 & 0.6485 & 0.7312 & 0.8551 & 0.5987 & 0.6960 & 0.8342 \\
    \midrule
    \multicolumn{1}{c|}{\multirow{15}[10]{*}{\parbox{1.5cm}{\centering All History GDP}}} & \multicolumn{1}{c|}{\multirow{3}[2]{*}{95-19}} & \multicolumn{1}{c|}{\multirow{3}[2]{*}{\texttimes }} & 8   & 1591: 168 & 1   & 0.5806 & 0.8245 & 0.9080 & 0.5560 & 0.8215 & 0.9064 & 0.5880 & 0.7861 & 0.8866 \\
        &     &     & 10  & 1549: 168 & 1   & 0.5563 & 0.7515 & 0.8669 & 0.7862 & 1.1409 & 1.0681 & 0.7071 & 0.8974 & 0.9473 \\
        &     &     & 12  & 1507: 168 & 1   & 0.5584 & 0.7547 & 0.8687 & 1.0564 & 1.7138 & 1.3091 & 0.7606 & 0.9580 & 0.9788 \\
\cmidrule{2-15}        & \multicolumn{1}{c|}{\multirow{12}[8]{*}{13-19}} & \multicolumn{1}{c|}{\multirow{3}[2]{*}{\texttimes }} & 8   & 336: 84 & 1   & 0.5253 & 0.6431 & 0.8020 & 0.8359 & 1.6756 & 1.2945 & 0.7185 & 1.5611 & 1.2494 \\
        &     &     & 10  & 294: 84 & 1   & 0.5180 & 0.5987 & 0.7738 & 0.7814 & 1.4338 & 1.1974 & 0.5903 & 0.6441 & 0.8026 \\
        &     &     & 12  & 252: 84 & 1   & 0.5850 & 0.7508 & 0.8665 & 0.7543 & 1.1243 & 1.0603 & 0.6681 & 1.0266 & 1.0132 \\
\cmidrule{3-15}        &     & \multicolumn{1}{c|}{\multirow{3}[2]{*}{\parbox{1cm}{\centering sum\textbackslash{} \\ mean\textbackslash{} \\ std}}} & 8   & 336: 84 & 4   & 0.5738 & 0.7142 & 0.8451 & 0.7426 & 1.2306 & 1.1093 & 0.7114 & 1.2258 & 1.1072 \\
        &     &     & 10  & 294: 84 & 4   & 0.6502 & 0.8551 & 0.9247 & 0.7444 & 1.5457 & 1.2433 & 0.6818 & 1.0567 & 1.0280 \\
        &     &     & 12  & 252: 84 & 4   & 0.7402 & 1.0441 & 1.0218 & 0.7022 & 1.0419 & 1.0207 & 0.6131 & 0.8448 & 0.9191 \\
\cmidrule{3-15}        &     & \multicolumn{1}{c|}{\multirow{3}[2]{*}{mean}} & 8   & 336: 84 & 2   & 0.5592 & 0.6959 & 0.8342 & 0.8074 & 1.3599 & 1.1662 & 0.6876 & 0.9716 & 0.9857 \\
        &     &     & 10  & 294: 84 & 2   & 0.6046 & 0.7821 & 0.8844 & 0.7608 & 1.5442 & 1.2426 & 0.6889 & 0.8934 & 0.9452 \\
        &     &     & 12  & 252: 84 & 2   & 0.6543 & 0.8960 & 0.9466 & 0.7555 & 1.1825 & 1.0874 & 0.7059 & 1.1972 & 1.0942 \\
\cmidrule{3-15}        &     & \multicolumn{1}{c|}{\multirow{3}[2]{*}{\parbox{1cm}{\centering every month mean}}} & 8   & 336: 84 & 4   & 0.5627 & 0.7151 & 0.8456 & 0.7112 & 1.1035 & 1.0505 & 0.6917 & 0.9704 & 0.9851 \\
        &     &     & 10  & 294: 84 & 4   & 0.6172 & 0.8069 & 0.8983 & 0.8365 & 1.8495 & 1.3599 & 0.6488 & 0.8087 & 0.8993 \\
        &     &     & 12  & 252: 84 & 4   & 0.6655 & 0.9066 & 0.9522 & 0.6530 & 0.9366 & 0.9678 & 0.8048 & 1.4220 & 1.1925 \\
    \bottomrule
    \end{tabular}%
}
  \label{tab:gdp-only-lstm}%
\end{table}%

Table \ref{tab:gdp-only-lstm} shows the prediction of the quarterly GDP growth results for linear regression and LSTM. We can see that for the first scenario, LSTM has comparable performance with linear regression, but linear regression is better than LSTM in the second scenario. Table \ref{tab:timesfm} is the performance of TimesFM by zero shot. TimesFM works for the univariable time series, hence the light data were not used. In the data type column, ``LSTM data" refers to using the data LSTM used as the input. ``Continuous data" refers to using all historical data before the label as the input. We can see that in all cases, TimesFM works best with all historical data before the label, but it is still worse than linear regression.

\begin{table}[htbp]
	\centering
	\caption{Prediction of the quarterly GDP growth by autoregression results for TimesFM. }
	\resizebox{\textwidth}{!}{%
		\renewcommand{\arraystretch}{1.2} 
		\begin{tabular}{>{\centering\arraybackslash}p{3cm}|>{\centering\arraybackslash}p{2cm}|>{\centering\arraybackslash}p{3cm}|>{\centering\arraybackslash}p{2cm}|>{\centering\arraybackslash}p{3cm}|>{\centering\arraybackslash}p{3cm}|>{\centering\arraybackslash}p{3cm}|>{\centering\arraybackslash}p{3cm}}
			\toprule
			\multirow{2}[4]{*}{Dataset} & \multirow{2}[4]{*}{Period} & \multirow{2}[4]{*}{Data type} & \multirow{2}[4]{*}{\parbox{1cm}{\centering Seq \\ Length}} & \multirow{2}[4]{*}{Train: Test} & \multicolumn{3}{c}{TimesFM} \\
			\cmidrule{6-8}
			&     &     &     &     & MAE & MSE & RMSE \\
			\midrule
			\multirow{8}[8]{*}{\parbox{3cm}{\centering GDP Filtered by Other Variables}}& \multirow{4}[4]{*}{95-19} & \multirow{3}[2]{*}{LSTM data} & 8   & 592: 108 & 1.8480 & 6.0209 & 2.4537 \\
			&     &     & 10  & 558: 104 & 1.8553 & 6.0676 & 2.4633 \\
			&     &     & 12  & 530: 100 & 1.8365 & 5.9708 & 2.4435 \\
			\cmidrule{3-8}        
			&     & Continuous data & -   & 787: 123 & 0.7691 & 1.5036 & 1.2262 \\
			\cmidrule{2-8}        
			& \multirow{4}[4]{*}{13-19} & \multirow{3}[2]{*}{LSTM data} & 8   & 202: 56 & 1.7839 & 5.6019 & 2.3668 \\
			&     &     & 10  & 170: 56 & 1.7838 & 5.5840 & 2.3630 \\
			&     &     & 12  & 143: 52 & 1.7797 & 5.5798 & 2.3622 \\
			\cmidrule{3-8}        
			&     & Continuous data & -   & 352: 60 & 0.6942 & 1.0255 & 1.0127 \\
			\midrule
			\multirow{8}[8]{*}{All History GDP } & \multirow{4}[4]{*}{95-19} & \multirow{3}[2]{*}{LSTM data} & 8   & 1591: 168 & 2.1024 & 7.5676 & 2.7509 \\
			&     &     & 10  & 1549: 168 & 2.1040 & 7.5600 & 2.7495 \\
			&     &     & 12  & 1507: 168 & 2.0954 & 7.4960 & 2.7379 \\
			\cmidrule{3-8}        
			&     & Continuous data & -   & 1759: 168 & 0.8045 & 1.5780 & 1.2562 \\
			\cmidrule{2-8}        
			& \multirow{4}[4]{*}{13-19} & \multirow{3}[2]{*}{LSTM data} & 8   & 336: 84 & 1.9659 & 6.5328 & 2.5559 \\
			&     &     & 10  & 294: 84 & 1.9771 & 6.5958 & 2.5682 \\
			&     &     & 12  & 252: 84 & 1.9757 & 6.5869 & 2.5665 \\
			\cmidrule{3-8}        
			&     & Continuous data & -   & 504: 84 & 0.6968 & 1.0522 & 1.0258 \\
			\bottomrule
		\end{tabular}%
	}
	\label{tab:timesfm}%
\end{table}

Table \ref{tab:gdppatch} shows the prediction of the quarterly GDP growth results for Time-LLM and PatchTST. We can see that generally Time-LLM and PatchTST are worse than linear regression. Tables \ref{tab:gdp-only-lstm}, \ref{tab:timesfm}, and \ref{tab:gdppatch} also show that the light data do not necessarily improve the prediction performance. 

\begin{table}[htbp]
  \centering
  \caption{Prediction of the quarterly GDP growth by autoregression results for Time-LLM and PatchTST}
    \resizebox{\textwidth}{!}{%
    \begin{tabular}{c|c|c|c|c|c|ccc|ccc|ccc|ccc}
    \toprule
    \multirow{2}[3]{*}{Dataset} & \multicolumn{1}{c|}{\multirow{2}[3]{*}{Period}} & \multicolumn{1}{c|}{\multirow{2}[3]{*}{Light}} & \multicolumn{1}{c|}{\multirow{2}[3]{*}{\parbox{1cm}{\centering Seq \\ Length}}} & \multicolumn{1}{c|}{\multirow{2}[3]{*}{Train: Test}} & \multirow{2}[3]{*}{Dims} & \multicolumn{3}{c|}{Time-LLM best Valid model} & \multicolumn{3}{c|}{Time-LLM Final Model} & \multicolumn{3}{c|}{PatchTST best Valid model} & \multicolumn{3}{c}{PatchTST Final Model} \\
\cmidrule{7-18}        &     &     & \multicolumn{1}{c|}{} & \multicolumn{1}{c|}{} &     & \multicolumn{1}{>{\centering\arraybackslash}p{3.75em}}{MAE} & \multicolumn{1}{>{\centering\arraybackslash}p{3.75em}}{MSE} & \multicolumn{1}{>{\centering\arraybackslash}p{3.75em}|}{RMSE} & \multicolumn{1}{>{\centering\arraybackslash}p{3.75em}}{MAE} & \multicolumn{1}{>{\centering\arraybackslash}p{3.75em}}{MSE} & \multicolumn{1}{>{\centering\arraybackslash}p{3.75em}|}{RMSE} & \multicolumn{1}{>{\centering\arraybackslash}p{3.75em}}{MAE} & \multicolumn{1}{>{\centering\arraybackslash}p{3.75em}}{MSE} & \multicolumn{1}{>{\centering\arraybackslash}p{3.75em}|}{RMSE} & \multicolumn{1}{>{\centering\arraybackslash}p{3.75em}}{MAE} & \multicolumn{1}{>{\centering\arraybackslash}p{3.75em}}{MSE} & \multicolumn{1}{>{\centering\arraybackslash}p{3.75em}}{RMSE} \\
\midrule
    \multicolumn{1}{c|}{\multirow{15}[9]{*}{\parbox{1.5cm}{\centering GDP Filtered by \\ Other Variables}}} & \multicolumn{1}{c|}{\multirow{3}[1]{*}{95-19}} & \multicolumn{1}{c|}{\multirow{3}[1]{*}{\texttimes}} & 8   & 592: 108 & 1   & 0.5656 & 0.6253 & 0.7908 & 0.5068 & 0.4826 & 0.6947 & 0.6529 & 0.8506 & 0.9223 & 0.5419 & 0.5810 & 0.7622 \\
        &     &     & 10  & 558: 104 & 1   & 0.5014 & 0.4901 & 0.7001 & 0.5844 & 0.6875 & 0.8291 & 0.4650 & 0.4320 & 0.6573 & 0.5352 & 0.5511 & 0.7424 \\
        &     &     & 12  & 530: 100 & 1   & 0.5770 & 0.6320 & 0.7950 & 0.6004 & 0.6459 & 0.8037 & 0.5000 & 0.4737 & 0.6883 & 0.5654 & 0.5716 & 0.7561 \\
\cmidrule{2-18}        & \multicolumn{1}{c|}{\multirow{12}[8]{*}{13-19}} & \multicolumn{1}{c|}{\multirow{3}[2]{*}{\texttimes}} & 8   & 202: 56 & 1   & 0.5616 & 0.6687 & 0.8178 & 0.4663 & 0.4075 & 0.6384 & 0.5304 & 0.5229 & 0.7231 & 0.7898 & 1.1882 & 1.0900 \\
        &     &     & 10  & 170: 56 & 1   & 0.5927 & 0.6877 & 0.8293 & 0.5648 & 0.6102 & 0.7812 & 0.5278 & 0.5260 & 0.7253 & 0.5780 & 0.6537 & 0.8085 \\
        &     &     & 12  & 143: 52 & 1   & 0.6471 & 0.6853 & 0.8278 & 0.6076 & 0.7627 & 0.8733 & 0.5937 & 0.6503 & 0.8064 & 0.5898 & 0.7461 & 0.8638 \\
\cmidrule{3-18}        &     & \multicolumn{1}{c|}{\multirow{3}[2]{*}{\parbox{1cm}{\centering sum\textbackslash{} \\ mean\textbackslash{} \\ std}}} & 8   & 202: 56 & 4   & 0.6701 & 0.7990 & 0.8939 & 0.6664 & 0.7286 & 0.8536 & 0.5681 & 0.6250 & 0.7906 & 0.6312 & 0.7144 & 0.8452 \\
        &     &     & 10  & 170: 56 & 4   & 0.6375 & 0.6775 & 0.8231 & 0.6821 & 0.8642 & 0.9296 & 0.6038 & 0.6476 & 0.8047 & 0.5303 & 0.5542 & 0.7445 \\
        &     &     & 12  & 143: 52 & 4   & 0.6838 & 0.8412 & 0.9172 & 0.6463 & 0.7792 & 0.8827 & 0.6049 & 0.7627 & 0.8733 & 0.5720 & 0.6313 & 0.7945 \\
\cmidrule{3-18}        &     & \multicolumn{1}{c|}{\multirow{3}[2]{*}{mean}} & 8   & 202: 56 & 2   & 0.6984 & 0.8362 & 0.9144 & 0.6392 & 0.7370 & 0.8585 & 0.6320 & 0.7035 & 0.8387 & 0.6920 & 0.8759 & 0.9359 \\
        &     &     & 10  & 170: 56 & 2   & 0.5757 & 0.6196 & 0.7872 & 0.7061 & 0.9379 & 0.9685 & 0.5613 & 0.5837 & 0.7640 & 0.4876 & 0.5046 & 0.7103 \\
        &     &     & 12  & 143: 52 & 2   & 0.6557 & 0.7636 & 0.8739 & 0.6022 & 0.6336 & 0.7960 & 0.6180 & 0.6248 & 0.7904 & 0.6208 & 0.7229 & 0.8503 \\
\cmidrule{3-18}        &     & \multicolumn{1}{c|}{\multirow{3}[2]{*}{\parbox{1cm}{\centering every month mean}}} & 8   & 202: 56 & 4   & 0.7604 & 0.9395 & 0.9693 & 0.6825 & 0.8792 & 0.9376 & 0.5546 & 0.6333 & 0.7958 & 0.6247 & 0.6600 & 0.8124 \\
        &     &     & 10  & 170: 56 & 4   & 0.6859 & 0.8064 & 0.8980 & 0.6368 & 0.7099 & 0.8426 & 0.4977 & 0.5187 & 0.7202 & 0.4242 & 0.3678 & 0.6065 \\
        &     &     & 12  & 143: 52 & 4   & 0.7347 & 0.9258 & 0.9622 & 0.6974 & 0.8916 & 0.9442 & 0.5200 & 0.5465 & 0.7393 & 0.6308 & 0.7184 & 0.8476 \\
    \midrule
    \multicolumn{1}{c|}{\multirow{15}[10]{*}{\parbox{1.5cm}{\centering All History GDP}}} & \multicolumn{1}{c|}{\multirow{3}[2]{*}{95-19}} & \multicolumn{1}{c|}{\multirow{3}[2]{*}{\texttimes}} & 8   & 1591: 168 & 1   & 0.7355 & 1.2720 & 1.1278 & 0.8102 & 1.7469 & 1.3217 & 0.6944 & 1.1046 & 1.0510 & 0.5795 & 0.9602 & 0.9799 \\
        &     &     & 10  & \multicolumn{1}{c|}{1549: 168} & 1   & 0.6416 & 0.8950 & 0.9461 & 0.6036 & 0.8385 & 0.9157 & 0.5546 & 0.7591 & 0.8713 & 0.6201 & 1.3994 & 1.1830 \\
        &     &     & 12  & 1507: 168 & 1   & 0.7678 & 1.4421 & 1.2009 & 0.7318 & 1.1586 & 1.0764 & 0.6695 & 1.2003 & 1.0956 & 0.5983 & 0.9708 & 0.9853 \\
\cmidrule{2-18}        & \multicolumn{1}{c|}{\multirow{12}[8]{*}{13-19}} & \multicolumn{1}{c|}{\multirow{3}[2]{*}{\texttimes}} & 8   & 336: 84 & 1   & 0.5734 & 0.8336 & 0.9130 & 0.7044 & 1.2048 & 1.0976 & 0.7936 & 1.4680 & 1.2116 & 0.9431 & 1.8501 & 1.3602 \\
        &     &     & 10  & 294: 84 & 1   & 0.6747 & 1.0961 & 1.0469 & 0.7213 & 1.0747 & 1.0367 & 0.8276 & 1.5585 & 1.2484 & 0.6521 & 0.8752 & 0.9355 \\
        &     &     & 12  & 252: 84 & 1   & 0.6859 & 1.0659 & 1.0324 & 0.6004 & 0.7841 & 0.8855 & 0.7709 & 1.0698 & 1.0343 & 0.5793 & 0.7322 & 0.8557 \\
\cmidrule{3-18}        &     & \multicolumn{1}{c|}{\multirow{3}[2]{*}{\parbox{1cm}{\centering sum\textbackslash{} \\ mean\textbackslash{} \\ std}}} & 8   & 336: 84 & 4   & 0.7125 & 1.0070 & 1.0035 & 0.6729 & 0.8736 & 0.9347 & 0.6417 & 0.9706 & 0.9852 & 0.6917 & 1.0041 & 1.0021 \\
        &     &     & 10  & 294: 84 & 4   & 0.7009 & 1.0885 & 1.0433 & 0.7578 & 1.1804 & 1.0865 & 0.5813 & 0.9046 & 0.9511 & 0.5805 & 0.6061 & 0.7785 \\
        &     &     & 12  & 252: 84 & 4   & 0.7734 & 1.3089 & 1.1441 & 0.8809 & 2.1906 & 1.4801 & 0.7958 & 1.9426 & 1.3938 & 0.5710 & 0.6508 & 0.8067 \\
\cmidrule{3-18}        &     & \multicolumn{1}{c|}{\multirow{3}[2]{*}{mean}} & 8   & 336: 84 & 2   & 0.8219 & 1.2894 & 1.1355 & 0.7860 & 1.2823 & 1.1324 & 0.7059 & 0.9449 & 0.9720 & 0.6891 & 1.3042 & 1.1420 \\
        &     &     & 10  & 294: 84 & 2   & 0.7704 & 1.1514 & 1.0730 & 0.6642 & 0.8862 & 0.9414 & 0.7400 & 1.3075 & 1.1435 & 0.6338 & 0.7524 & 0.8674 \\
        &     &     & 12  & 252: 84 & 2   & 0.6897 & 1.0560 & 1.0276 & 0.7488 & 1.3487 & 1.1613 & 0.7747 & 1.3590 & 1.1658 & 0.6257 & 0.8307 & 0.9114 \\
\cmidrule{3-18}        &     & \multicolumn{1}{c|}{\multirow{3}[2]{*}{\parbox{1cm}{\centering every month mean}}} & 8   & 336: 84 & 4   & 0.8889 & 1.6304 & 1.2769 & 0.8151 & 1.3906 & 1.1792 & 0.6544 & 1.0982 & 1.0480 & 0.5748 & 0.8403 & 0.9167 \\
        &     &     & 10  & 294: 84 & 4   & 0.8912 & 1.6782 & 1.2955 & 0.8107 & 1.5971 & 1.2638 & 0.6516 & 0.9422 & 0.9707 & 0.7434 & 1.4285 & 1.1952 \\
        &     &     & 12  & 252: 84 & 4   & 0.7714 & 1.3029 & 1.1414 & 0.8588 & 1.5580 & 1.2482 & 0.5877 & 0.7816 & 0.8841 & 0.6420 & 0.8649 & 0.9300 \\
    \bottomrule
    \end{tabular}%
    }
  \label{tab:gdppatch}%
\end{table}%

\section{Prediction of the quarterly GDP growth with multi-indicator data}\label{sec:premulgdp}

In Section \ref{sec:pregdp}, we predict the quarterly GDP growth by autoregression, but models there could not characterize the impacts of previous economic indicators on GDP growth. Hence, in this section, we predict the quarterly GDP growth by previous economic indicators, i.e., predict $y^t$ by $(z^{t-h}, ..., z^{t-1})$, where $z^i$ is defined in (\ref{eq:zt}). We use four models in this section, i.e., linear regression, LSTM, Time-LLM \citep{jin2023time}, and PatchTST \citep{nie2022time}. For linear regression, LSTM, and PatchTST, we actually predict $z^t$ by vector autoregression. Since the goal is predicting $y^t$, hence we give more weight to the GDP growth in the loss function for LSTM and PatchTST. The loss function actually has the following form 
\begin{equation}
f_{loss} = \sum_{i=1}^n (x_i^t - {\hat x}_i^t)^2 + W_{GDP} (y^t - {\hat y}^t)^2, 
\end{equation}
where $({\hat x}_1^t, ..., {\hat x}_i^t, {\hat y}^t)$ is the prediction for $z^t$ and $W_{GDP} > 0$ is the weight for the GDP growth which is a hyperparameter. For the validation loss for LSTM, Time-LLM, and PatchTST, we only calculate the GDP growth part.

\begin{table}[htbp]
  \centering
  \caption{Prediction of the quarterly GDP growth with multi-indicator data results for linear regression and LSTM.} 
  \resizebox{\textwidth}{!}{%
    \renewcommand{\arraystretch}{1.2} 
    \begin{tabular}{>{\centering\arraybackslash}p{2cm}|>{\centering\arraybackslash}p{1cm}|>{\centering\arraybackslash}p{1cm}|c|c|>{\centering\arraybackslash}p{1cm}|ccc|ccc|ccc}
    \toprule
    \multirow{2}[3]{*}{Dataset} & \multicolumn{1}{c|}{\multirow{2}[3]{*}{Period}} & \multicolumn{1}{c|}{\multirow{2}[3]{*}{Light}} & \multicolumn{1}{>{\centering\arraybackslash}p{1cm}|}{\multirow{2}[3]{*}{{\parbox{1cm}{\centering Seq Length}}}} & \multicolumn{1}{c|}{\multirow{2}[3]{*}{Train: Test}} & \multirow{2}[3]{*}{Dims} & \multicolumn{3}{c|}{Linear Regression} & \multicolumn{3}{c|}{LSTM Best Valid Model} & \multicolumn{3}{c}{LSTM Final Model} \\
    \cmidrule{7-15}
    &     &     & \multicolumn{1}{c|}{} & \multicolumn{1}{c|}{} &     & \multicolumn{1}{>{\centering\arraybackslash}p{2cm}}{MAE} & \multicolumn{1}{>{\centering\arraybackslash}p{2cm}}{MSE} & \multicolumn{1}{>{\centering\arraybackslash}p{2cm}|}{RMSE} & \multicolumn{1}{>{\centering\arraybackslash}p{2cm}}{MAE} & \multicolumn{1}{>{\centering\arraybackslash}p{2cm}}{MSE} & \multicolumn{1}{>{\centering\arraybackslash}p{2cm}|}{RMSE} & \multicolumn{1}{>{\centering\arraybackslash}p{2cm}}{MAE} & \multicolumn{1}{>{\centering\arraybackslash}p{2cm}}{MSE} & \multicolumn{1}{>{\centering\arraybackslash}p{2cm}}{RMSE} \\
    \midrule
    \multirow{15}[9]{*}{\parbox{2cm}{\centering GDP with Other Variables }} & \multirow{3}[1]{*}{95-19} & \multirow{3}[1]{*}{\texttimes} & 8   & 592: 108 & 21  & 0.6626 & 0.7761 & 0.8810 & 0.7845 & 0.9352 & 0.9670 & 0.6623 & 0.7122 & 0.8439 \\
        &     &     & 10  & 558: 104 & 21  & 0.9189 & 1.8856 & 1.3732 & 0.8484 & 1.1802 & 1.0864 & 0.7097 & 0.8586 & 0.9266 \\
        &     &     & 12  & 530: 100 & 21  & 0.9225 & 1.6033 & 1.2662 & 0.7122 & 0.8166 & 0.9037 & 0.7659 & 0.9043 & 0.9510 \\
    \cmidrule{2-15}   
    & \multirow{12}[8]{*}{13-19} & \multirow{3}[2]{*}{\texttimes} & 8   & 202: 56 & 21  & 3.6646 & 34.8454 & 5.9030 & 0.6869 & 0.8769 & 0.9364 & 0.6958 & 0.8727 & 0.9342 \\
        &     &     & 10  & 170: 56 & 21  & 2.5218 & 15.1833 & 3.8966 & 0.7555 & 1.0397 & 1.0196 & 0.7910 & 0.9904 & 0.9952 \\
        &     &     & 12  & 143: 52 & 21  & 1.2476 & 2.8113 & 1.6767 & 1.0640 & 1.7409 & 1.3194 & 0.9097 & 1.2224 & 1.1056 \\
    \cmidrule{3-15}        
    &     & \multirow{3}[2]{*}{\parbox{1cm}{\centering sum\textbackslash{} \\ mean\textbackslash{} \\ std}} & 8   & 202: 56 & 24  & 5.0465 & 105.0620 & 10.2500 & 0.7447 & 1.0500 & 1.0247 & 0.7443 & 1.2163 & 1.1029 \\
        &     &     & 10  & 170: 56 & 24  & 1.6424 & 4.0354 & 2.0088 & 0.7717 & 0.9947 & 0.9974 & 0.8862 & 1.2894 & 1.1355 \\
        &     &     & 12  & 143: 52 & 24  & 1.1717 & 2.1186 & 1.4555 & 0.8111 & 1.1020 & 1.0497 & 0.7885 & 1.1107 & 1.0539 \\
    \cmidrule{3-15}        
    &     & \multirow{3}[2]{*}{mean} & 8   & 202: 56 & 22  & 3.1707 & 37.9830 & 6.1630 & 0.5667 & 0.7400 & 0.8602 & 0.7776 & 1.1846 & 1.0884 \\
        &     &     & 10  & 170: 56 & 22  & 1.8926 & 7.0782 & 2.6605 & 0.6640 & 0.7797 & 0.8830 & 0.7645 & 0.9739 & 0.9869 \\
        &     &     & 12  & 143: 52 & 22  & 1.2036 & 2.5812 & 1.6066 & 0.8069 & 1.0748 & 1.0367 & 0.8759 & 1.1421 & 1.0687 \\
    \cmidrule{3-15}        
    &     & \multirow{3}[2]{*}{\parbox{1cm}{\centering every month mean}} & 8   & 202: 56 & 24  & 3.4789 & 35.8779 & 5.9898 & 0.7356 & 0.9659 & 0.9828 & 0.6097 & 0.7380 & 0.8591 \\
        &     &     & 10  & 170: 56 & 24  & 1.8478 & 6.1222 & 2.4743 & 0.7755 & 1.1304 & 1.0632 & 0.6910 & 0.8486 & 0.9212 \\
        &     &     & 12  & 143: 52 & 24  & 1.0143 & 1.7284 & 1.3147 & 0.7108 & 1.0255 & 1.0127 & 0.9846 & 1.3968 & 1.1818 \\
    \bottomrule
    \end{tabular}%
  }
  \label{tab:gdpmul}%
\end{table}%

Table \ref{tab:gdpmul} shows the prediction of the quarterly GDP growth with multi-indicator data results for linear regression and LSTM. It indicates that LSTM is generally better than linear regression in the multi-indicator case.

\begin{table}[htbp]
  \centering
  \caption{Prediction of the quarterly GDP growth with multi-indicator data results for Time-LLM and PatchTST.}
  \resizebox{\textwidth}{!}{%
    \begin{tabular}{c|c|c|>{\centering\arraybackslash}p{1cm}|c|c|ccc|ccc|ccc|ccc}
    \toprule
    \multirow{2}[3]{*}{Dataset} & \multicolumn{1}{c|}{\multirow{2}[3]{*}{Period}} & \multicolumn{1}{c|}{\multirow{2}[3]{*}{Light}} & \multicolumn{1}{c|}{\multirow{2}[3]{*}{\parbox{1cm}{\centering Seq Length}}} & \multicolumn{1}{c|}{\multirow{2}[3]{*}{Train: Test}} & \multirow{2}[3]{*}{Dims} & \multicolumn{3}{c|}{Time-LLM best Valid model} & \multicolumn{3}{c|}{Time-LLM Final Model} & \multicolumn{3}{c|}{PatchTST best Valid model} & \multicolumn{3}{c}{PatchTST Final Model} \\
\cmidrule{7-18}        &     &     & \multicolumn{1}{c|}{} & \multicolumn{1}{c|}{} &     & \multicolumn{1}{c}{MAE} & \multicolumn{1}{c}{MSE} & \multicolumn{1}{c|}{RMSE} & \multicolumn{1}{c}{MAE} & \multicolumn{1}{c}{MSE} & \multicolumn{1}{c|}{RMSE} & \multicolumn{1}{c}{MAE} & \multicolumn{1}{c}{MSE} & \multicolumn{1}{c|}{RMSE} & \multicolumn{1}{c}{MAE} & \multicolumn{1}{c}{MSE} & \multicolumn{1}{c}{RMSE} \\
\midrule
\multicolumn{1}{c|}{\multirow{15}[9]{*}{\parbox{1cm}{\centering GDP with \\ Other Variables}}} & \multicolumn{1}{c|}{\multirow{3}[1]{*}{95-19}} & \multicolumn{1}{c|}{\multirow{3}[1]{*}{\texttimes}} & 8   & 592: 108 & 21  & 0.5842 & 0.6392 & 0.7995 & 0.5348 & 0.5681 & 0.7537 & 0.6052 & 0.7117 & 0.8436 & 0.6854 & 0.8860 & 0.9413 \\
        &     &     & 10  & 558: 104 & 21  & 0.5258 & 0.4999 & 0.7071 & 0.5392 & 0.6115 & 0.7820 & 0.4884 & 0.5130 & 0.7163 & 0.4928 & 0.4425 & 0.6652 \\
        &     &     & 12  & 530: 100 & 21  & 0.4724 & 0.4376 & 0.6615 & 0.5421 & 0.5605 & 0.7487 & 0.5005 & 0.4671 & 0.6834 & 0.5386 & 0.5957 & 0.7718 \\
\cmidrule{2-18}        & \multicolumn{1}{c|}{\multirow{12}[8]{*}{13-19}} & \multicolumn{1}{c|}{\multirow{3}[2]{*}{\texttimes}} & 8   & 202: 56 & 21  & 0.4988 & 0.5435 & 0.7372 & 0.4992 & 0.4931 & 0.7022 & 0.5939 & 0.6322 & 0.7951 & 0.5125 & 0.5102 & 0.7143 \\
        &     &     & 10  & 170: 56 & 21  & 0.4932 & 0.5202 & 0.7213 & 0.5305 & 0.5865 & 0.7658 & 0.9690 & 1.6141 & 1.2705 & 0.4497 & 0.4353 & 0.6597 \\
        &     &     & 12  & 143: 52 & 21  & 0.5736 & 0.5998 & 0.7745 & 0.5992 & 0.6919 & 0.8318 & 0.5285 & 0.5526 & 0.7434 & 0.6126 & 0.7758 & 0.8808 \\
\cmidrule{3-18}        &     & \multicolumn{1}{c|}{\multirow{3}[2]{*}{\parbox{1cm}{\centering sum\textbackslash{} \\ mean\textbackslash{} \\ std}}} & 8   & 202: 56 & 24  & 0.5469 & 0.6442 & 0.8026 & 0.5545 & 0.5648 & 0.7515 & 0.5186 & 0.5373 & 0.7330 & 0.5265 & 0.5301 & 0.7281 \\
        &     &     & 10  & 170: 56 & 24  & 0.5612 & 0.6347 & 0.7967 & 0.5651 & 0.5349 & 0.7314 & 0.5064 & 0.4725 & 0.6874 & 0.4697 & 0.4474 & 0.6689 \\
        &     &     & 12  & 143: 52 & 24  & 0.5434 & 0.5756 & 0.7587 & 0.5749 & 0.5836 & 0.7639 & 0.5964 & 0.6630 & 0.8142 & 0.5264 & 0.5974 & 0.7729 \\
\cmidrule{3-18}        &     & \multicolumn{1}{c|}{\multirow{3}[2]{*}{mean}} & 8   & 202: 56 & 22  & 0.5544 & 0.5670 & 0.7530 & 0.5527 & 0.6226 & 0.7891 & 0.6648 & 0.7893 & 0.8884 & 0.5149 & 0.4606 & 0.6787 \\
        &     &     & 10  & 170: 56 & 22  & 0.5569 & 0.6488 & 0.8055 & 0.5914 & 0.6663 & 0.8163 & 0.5345 & 0.5690 & 0.7543 & 0.4831 & 0.5147 & 0.7174 \\
        &     &     & 12  & 143: 52 & 22  & 0.6126 & 0.7070 & 0.8408 & 0.6450 & 0.7164 & 0.8464 & 0.5962 & 0.6543 & 0.8089 & 0.5830 & 0.5950 & 0.7713 \\
\cmidrule{3-18}        &     & \multicolumn{1}{c|}{\multirow{3}[2]{*}{\parbox{1cm}{\centering every month mean}}} & 8   & 202: 56 & 24  & 0.6115 & 0.7286 & 0.8536 & 0.5841 & 0.6204 & 0.7877 & 0.5996 & 0.6683 & 0.8175 & 0.5737 & 0.5634 & 0.7506 \\
        &     &     & 10  & 170: 56 & 24  & 0.6017 & 0.6898 & 0.8305 & 0.5902 & 0.6717 & 0.8196 & 0.5696 & 0.5047 & 0.7104 & 0.4942 & 0.4035 & 0.6352 \\
        &     &     & 12  & 143: 52 & 24  & 0.6826 & 0.8536 & 0.9239 & 0.6176 & 0.6912 & 0.8314 & 0.6162 & 0.6780 & 0.8234 & 0.7491 & 0.9456 & 0.9724 \\
    \bottomrule
    \end{tabular}%
    }
  \label{tab:gdpmulpatch}%
\end{table}%

Table \ref{tab:gdpmulpatch} shows the prediction of the quarterly GDP growth with multi-indicator data results for Time-LLM and PatchTST. They are actually comparable and are both better than LSTM generally. It should be noticed that, Time-LLM deals with each channel independently and PatchTST has an individual head for each channel and also deal with each channel independently, hence they actually could not characterize the impacts of economic indicators during the inference phase. Tables \ref{tab:gdpmul} and \ref{tab:gdpmulpatch} also indicate that the light data do not necessarily improve the prediction performance.

	\bibliographystyle{unsrtnat}
	\bibliography{gdp.bib}

	\appendix

	\part*{Appendix}


	\section{Descriptions used by RT and Time-LLM}

\begin{table}[htbp]
  \centering
  \caption{Descriptions used by RT}
  \resizebox{\textwidth}{!}{ 
    \begin{tabular}{|p{21.635em}|p{48.955em}|}
    \toprule
    \textbf{Variable} & \textbf{Description} \\
    \midrule
    Rural population growth (annual \%) & The index 'Rural population growth (annual \%)' measures the annual percentage increase or decrease in the rural population of a country. In this year, the index is {value}. \\
    \midrule
    General government final consumption expenditure (annual \% growth) & The index 'General government final consumption expenditure (annual \% growth)' measures the annual percentage increase or decrease in government spending on goods and services that are used for providing public services. In this year, the index is {value}. \\
    \midrule
    Consumer price index (2010 = 100) & The index 'Consumer price index (2010 = 100)' measures the average change over time in the prices paid by consumers for a basket of goods and services, with the base year set to 2010. This indicator is essential for tracking inflation, assessing the cost of living, and guiding monetary policy decisions. In this year, the index is {value}. \\
    \midrule
    Exports of goods and services (annual \% growth) & The index 'Exports of goods and services (annual \% growth)' measures the annual percentage increase or decrease in the value of a country's exports of goods and services. In this year, the index is {value}. \\
    \midrule
    Urban population growth (annual \%) & The index 'Urban population growth (annual \%)' measures the annual percentage increase or decrease in the population residing in urban areas. In this year, the index is {value}. \\
    \midrule
    Population growth (annual \%) & The index 'Population growth (annual \%)' measures the annual percentage increase or decrease in the total population of a country. In this year, the index is {value}. \\
    \midrule
    Inflation, GDP deflator (annual \%) & The index 'Inflation, GDP deflator (annual \%)' measures the annual percentage change in the price level of all new, domestically produced, final goods and services in an economy. In this year, the index is {value}. \\
    \midrule
    Imports of goods and services (annual \% growth) & The index 'Imports of goods and services (annual \% growth)' measures the annual percentage increase or decrease in the value of a country's imports of goods and services. In this year, the index is {value}. \\
    \midrule
    Final consumption expenditure (annual \% growth) & The index 'Final consumption expenditure (annual \% growth)' measures the annual percentage change in the total value of all goods and services consumed by households and government. In this year, the index is {value}. \\
    \midrule
    Unemployment, total (\% of total labor force) (national estimate) & The index 'Unemployment, total (\% of total labor force) (national estimate)' measures the percentage of the total labor force that is unemployed and actively seeking employment, based on national estimates. In this year, the index is {value}. \\
    \midrule
    Inflation, consumer prices (annual \%) & The index 'Inflation, consumer prices (annual \%)' measures the annual percentage change in the average level of prices for consumer goods and services. In this year, the index is {value}. \\
    \midrule
    Gross fixed capital formation (annual \% growth) & The index 'Gross fixed capital formation (annual \% growth)' measures the annual percentage increase or decrease in investment in fixed assets such as buildings, machinery, and infrastructure. In this year, the index is {value}. \\
    \midrule
    Households and NPISHs Final consumption expenditure (annual \% growth) & The index 'Households and NPISHs Final consumption expenditure (annual \% growth)' measures the annual percentage change in the spending by households and Non-Profit Institutions Serving Households (NPISHs) on goods and services. In this year, the index is {value}. \\
    \midrule
    Nighttime light (sum\textbackslash{}mean\textbackslash{}std) & Nighttime light remote sensing data refers to the use of remote sensing technology to capture the distribution of lights on Earth at night. It can effectively reflect the spatial distribution of human activities and is therefore commonly used in remote sensing inversion of various socio-economic data. In this data, each pixel represents the light intensity of a geographical area of 500 meters by 500 meters. In this year, the total sum of light intensity of all pixels occupied by the country or region is {sum}, the average is {mean}, the standard deviation is {std}. \\
    \midrule
    Nighttime light (every month mean) & Nighttime light remote sensing data refers to the use of remote sensing technology to capture the distribution of lights on Earth at night. In this data, each pixel represents the light intensity of a geographical area of 500 meters by 500 meters. In the {month\_num}th month of this year, the average of light intensity of all pixels occupied by the country or region is {mean}. \\
    \bottomrule
    \end{tabular}%
    }
  \label{tab:desRT}%
\end{table}%

\begin{table}[htbp]
  \centering
  \caption{Descriptions used by Time-LLM}
  \resizebox{\textwidth}{!}{ 
    \begin{tabular}{|p{18.09em}|p{37.955em}|}
    \toprule
    \textbf{Dataset Type} & \textbf{Description} \\
    \midrule
    Multi-indicators & Gross Domestic Product (GDP) is a measure of the total monetary value of all finished goods and services produced within a country’s borders in a specific time period. This dataset consists some economic indicators, such as 'Export Value', 'Industrial Added Value', 'Stock Market Capitalization', 'Balance of Payments - Financial Account Balance', 'Net International Investment Position', 'Import Value', 'Nominal Effective Exchange Rate', 'Retail Sales', 'CPI (Consumer Price Index)', 'Unemployment Rate' and 'Central Bank Policy Rate', etc. \\
    \midrule
    Only gdp & Gross Domestic Product (GDP) is a measure of the total monetary value of all finished goods and services produced within a country’s borders in a specific time period. This dataset consists only GDP. \\
    \midrule
    Gdp and light & Gross Domestic Product (GDP) is a measure of the total monetary value of all finished goods and services produced within a country’s borders in a specific time period. This dataset consists GDP and some Nighttime light remote sensing data, which refers to the use of remote sensing technology to capture the distribution of lights on Earth at night. \\
    \bottomrule
    \end{tabular}%
  }
  \label{tab:destime-llm}%
\end{table}%

\end{document}